\documentclass[sigconf]{acmart}

\AtBeginDocument{%
  \providecommand\BibTeX{{%
    \normalfont B\kern-0.5em{\scshape i\kern-0.25em b}\kern-0.8em\TeX}}}

\usepackage[portuguese]{babel}
\usepackage[utf8]{inputenc}
\usepackage{graphicx}

\begin{document}

\title{Avaliação de desempenho de métodos de snapshot para aquecer o cold-start de funções como serviço}

\author{Paulo Silva}
\email{paulo.felipe.silva@ccc.ufcg.edu.br}
\orcid{1234-5678-9012}
\authornotemark[1]
\affiliation{%
  \institution{Universidade Federal de Campina Grande}
  \streetaddress{R. Aprígio Veloso, 882 - Universitário}
  \city{Campina Grande}
  \state{Paraíba}
  \country{Brasil}
  \postcode{58428-830}
}

\author{Thiago Emmanuel Pereira}
\email{temmanuel@computacao.ufcg.edu.br}
\affiliation{%
  \institution{Universidade Federal de Campina Grande}
  \streetaddress{R. Aprígio Veloso, 882 - Universitário}
  \city{Campina Grande}
  \state{Paraíba}
  \country{Brasil}
  \postcode{58428-830}
}

\renewcommand{\shortauthors}{Silva and Pereira.}

\begin{abstract}
  O modelo de computação \textit{serverless} fortaleceu a tendência da computação em nuvem de tornar transparente o gerenciamento da infraestrutura. Ao simplificar o gerenciamento, o modelo \textit{serverless} deixa a responsabilidade de implantação e escalonamento para a plataforma. Aliado a isso, com um modelo de cobrança que considera somente o tempo despendido com a execução de requisições, há um forte incentivo para o uso eficiente dos recursos. Essa busca por eficiência, traz à tona o problema de \textit{cold-start}, que se configura como um atraso na execução de aplicações \textit{serverless}. Dentre as soluções propostas para lidar com o \textit{cold-start}, se destacam as baseadas no método de \textit{snapshot}. Apesar da exploração desse método, existe uma carência de trabalhos que avaliam os \textit{trade-offs} de cada proposta. Nessa direção, este trabalho compara duas estratégias de mitigação do \textit{cold-start}: \textit{Prebaking} e \textit{SEUSS}. Avaliamos o desempenho das estratégias experimentalmente com funções de diferentes níveis de complexidade: NoOp, uma função que converte \textit{markdown} para HTML, e uma que carrega 41 MB de dependências. Resultados preliminares indicam que \textit{Prebaking} apresentou desempenho 33\% e 25\% superior para inicializar \textit{NoOp} e \textit{Markdown}, respectivamente e processou a primeira requisição de \textit{Markdown} com um tempo 69\% inferior ao \textit{SEUSS}.
\end{abstract}

\maketitle

%PARA O TCC: Use esta nota de rodapé  na introdução para o artigo final

\section{Introdução}
\label{section:introduction}
Computação \textit{serverless} é um novo tipo de oferta de computação em nuvem, que retira a responsabilidade dos desenvolvedores de gerenciar a infraestrutura para provisionar a execução de aplicações. Serviços \textit{serverless} possuem um modelo de cobrança que contabiliza apenas os recursos utilizados durante a execução das aplicações. A oferta mais comum de \textit{serverless} utiliza o modelo de computação \textit{FaaS} (Função como Serviço).

Plataformas de Função como Serviço definem que as unidades de implantação e gerência das aplicações são funções. Dessa forma, os componentes arquiteturais de uma aplicação \textit{FaaS} devem ser abstraídos em uma função. Esse princípio força os desenvolvedores a adotarem uma metodologia de programação simples e coesa.

\textit{Serverless} vêm ganhando popularidade \cite{passwater2018serverless}. Esse aumento é refletido no lançamento de várias plataformas nos últimos anos \cite{amazonwebservices, googlecloud, microsoftazure, ibmcloud, huaweicloud}. Uma das razões da popularidade da computação \textit{serverless} é que os desenvolvedores podem focar na implementação da lógica de negócio, deixando de lado detalhes de implantação. O ganho dessa estratégia é a aceleração do tempo de colocação no mercado de serviços web com a garantia de escalonamento de acordo com a demanda de utilização. 

Um outro atrativo do modelo \textit{serverless}, é o reforço da tendência de cobrança \textit{pay-as-you-go}, inaugurada pelas ofertas de \textit{IaaS} (Infraestrutura como Serviço). Por exemplo, se uma aplicação \textit{FaaS} estiver disponível durante um dia inteiro, mas for requisitada apenas uma vez, durante 1s, a plataforma deve faturar apenas o 1s de computação efetivamente realizada.

Uma vez que os usuários pagam somente devido à execução das funções, há um incentivo natural para que o provedor não mantenha recursos computacionais alocados desnecessariamente. Essa prática pode levar ao conhecido problema de \textit{cold-start}, que se configura como um atraso na execução de funções quando não existe nenhuma instância disponível para executá-la.

No momento do \textit{cold-start}, o usuário terá que esperar a plataforma realizar todo o processo de criação de uma instância para executar a função. Entre as etapas envolvidas para criação de uma instância de uma função, temos:

\begin{enumerate}
    \item a criação do ambiente de execução isolada;
    \item a criação do processo e da \textit{runtime} que irão gerenciar a execução da função;
    \item e por fim, as etapas de \textit{startup} específicas da aplicação como a inicialização de
serviços HTTP, e carregamento do código fonte da função e suas dependências.
\end{enumerate}

Normalmente as plataformas \textit{serverless} isolam a execução das funções utilizando contêineres ou VMs \cite{wang2018peeking, jonas2019cloud}. O \textit{startup} desses componentes parece ser a etapa mais longa de inicialização \cite{agache2020firecracker, cadden2020seuss}. Porém, \textit{Prebaking} \cite{silva2020prebaking} mostrou que quanto mais complexa a aplicação, mais o \textit{cold-start} das etapas 2 e 3 contribuem significativamente no atraso da execução das funções. Algumas soluções foram propostas para lidarem com as etapas 2 e 3 \cite{oakes2018sock, boucher2018putting}. Dentre estas soluções se destacam as que se baseiam em métodos de \textit{snapshot} \cite{silva2020prebaking, cadden2020seuss, wang2019replayable, du2020catalyzer}, pois tendem a não realizar grandes mudanças arquitetônicas nas plataformas já existentes, e não aumentar a complexidade de operação.

Diversos trabalhos já foram publicados demonstrando a eficácia do método de \textit{snapshot} para acelerar o \textit{cold-start}. Em particular, apesar de se basearem no mesmo método, a técnica de \textit{Prebaking} e o método \textit{SEUSS} utilizam estratégias distintas para alcançar o mesmo objetivo. Cada estratégia possui suas vantagens e desvantagens e não existe nenhum trabalho que discute os cenários onde cada estratégia é mais adequada.

Uma vez que o método de \textit{snapshot} ainda não é amplamente utilizado por plataformas \textit{serverless}, expor os \textit{trade-offs} de cada estratégia pode ajudar as plataformas a decidirem qual método pode ser mais adequado aos seus requisitos. Dessa forma, este trabalho compara o desempenho da técnica de \textit{Prebaking} e do método \textit{SEUSS} para acelerar o \textit{startup} de aplicações \textit{serverless}. A comparação foi realizada via experimentos de medição do tempo de inicialização e o tempo de processamento de funções \textit{serverless}. As funções avaliadas foram: 1) \textit{NoOp} que não faz nada; 2) \textit{Markdown} que converte \textit{markdown} para HTML; e 3) \textit{Big} que carrega 41 MB de dependências externas.

Os resultados indicam que \textit{Prebaking} teve um desempenho superior em relação ao \textit{SEUSS} para acelerar a inicialização das funções \textit{NoOp} e \textit{Markdown}. Para a função \textit{NoOp} o tempo de inicialização mediano do \textit{SEUSS} ficou entre 12 ms e 14 ms, enquanto \textit{Prebaking} foi de 8 ms, apresentando uma diferença percentual de 33\% e 45\%. Para \textit{Markdown}, uma função um pouco mais complexa, o tempo de inicialização mediano do \textit{SEUSS} ficou entre 12 ms e 13 ms, representando uma diferença entre 25\% e 30\% em relação à \textit{Prebaking}.

Além disso, foi possível observar a importância da análise do impacto do \textit{cold-start} no tempo de processamento da primeira requisição. Essa análise permite mensurar possíveis penalidades no processamento das requisições, assim como medir a eficácia de mecanismos de \textit{warmup} para reduzir este efeito. O tempo de serviço da primeira requisição processada pela função \textit{Markdown} com o método \textit{SEUSS} apresentou uma penalidade de 6 ms, um percentual de 9,6\%. Por outro lado, a mesma função com a técnica de \textit{Prebaking}, que adota um mecanismo de \textit{warmup}, apresentou uma redução de 11 ms no tempo de serviço da primeira requisição, representando uma redução percentual de 34\%.

\section{Contextualização}

Salvar e restaurar o estado de um programa ou processo é um mecanismo comumente empregado para promover a tolerância à falhas de sistemas. Dessa forma, caso ocorra uma falha na execução de um programa, o estado do processo pode ser recuperado utilizando a versão do processo mantida em armazenamento estável, chamada de \textit{snapshot} \cite{koo1987checkpointing}. Restaurar o processo evita recomputar todas as operações necessárias para que o programa volte ao estado anterior a falha ou interrupção.

Como o \textit{cold-start} se configura como um atraso na inicialização de uma aplicação, a técnica de salvar e restaurar o estado de um programa pode ser utilizada para evitar a execução de etapas de inicialização. Com isso, as plataformas \textit{serverless} podem criar \textit{snapshots} das aplicações já em um estado pronto para processar requisições, e restaurar esses \textit{snapshots} sempre que for necessário criar uma nova instância da aplicação. 

\subsection{Prebaking}

Uma das soluções propostas, que emprega o método de \textit{snapshot} para reduzir o efeito do \textit{cold-start}, é a técnica de \textit{Prebaking} \cite{silva2020prebaking}. A técnica define que antes da efetiva implantação na plataforma, a aplicação deve ser executada de maneira offline para a geração do seu \textit{snapshot}. E antes desse processo, a plataforma deve forçar a realização de diversos passos de inicialização, incluindo: carregar e compilar o código da aplicação junto com suas dependências. Os resultados reportados pela técnica demonstraram que a definição do momento da geração do \textit{snapshot} é um fator que também influencia no \textit{cold-start}.

Com a execução de todos os passos de inicialização e com a aplicação pronta para processar requisições, \textit{Prebaking} utiliza o CRIU\footnote{\url{https://www.criu.org/Main_Page}} para gerar o \textit{snapshot} da aplicação. O CRIU é uma ferramenta que permite, em espaço de usuário, salvar o estado de um processo do sistema operacional Linux em armazenamento estável. 

Com o estado da aplicação em disco, a técnica pressupõe que a plataforma poderá distribuir o \textit{snapshot}, ao invés do código da aplicação, para ser utilizado como um artefato de implantação de novas instâncias. Assim, múltiplas réplicas de uma aplicação podem ser criadas a partir do mesmo \textit{snapshot}. Ou seja, sempre que for necessário criar uma nova réplica da aplicação, a plataforma pode carregar o \textit{snapshot}, caso já não esteja disponível, e em seguida restaurar o processo da aplicação já pronto para processar requisições dos clientes.

A avaliação da técnica demonstrou que o método de \textit{snapshot} tem o potencial de acelerar o \textit{cold-start} de aplicações \textit{serverless}, em particular de plataformas \textit{FaaS}. Os resultados demonstraram ganhos que variam de 40\% a 71\% para aplicações básicas do modelo \textit{serverless}. E para aplicações mais complexas, o ganho pode ser ainda maior. Por exemplo, em aplicações que carregam uma grande quantidade de dependências, houve uma redução no tempo de inicialização de x19.

\subsection{SEUSS}

Outros trabalhos também já demonstraram a eficácia do método de \textit{snapshot}. O \textit{SEUSS} (\textit{Serverless Execution via Unikernel SnapShot}) é um método que tem como objetivo acelerar a implantação e diminuir o \textit{memory-footprint} de aplicações \textit{serverless} \cite{cadden2020seuss}. Apesar do desempenho do \textit{cold-start} do método para funções mais complexas ainda ser desconhecido, a estratégia alcançou uma latência de \textit{cold-start} de 7,5 ms no pior caso para uma função \textit{NoOp}. 

Um dos motivos para a redução considerável no \textit{cold-start}, foi a adoção de unikernels para isolar a execução das aplicações. Unikernels são kernels especializados que oferecem componentes do sistema operacional como bibliotecas \cite{raza2019unikernels}. Em um unikernel, a aplicação e o kernel são combinados no mesmo espaço de endereçamento de memória, criando um executável que pode ser inicializado diretamente em hardwares virtualizados \cite{madhavapeddy2014unikernels}.

Com unikernels a aplicação pode importar apenas as funcionalidades realmente necessárias do kernel. Aplicações na nuvem podem se beneficiar com essa estratégia removendo diversas camadas de software desnecessárias, que foram pensadas para sistemas operacionais de propósito geral. A redução da quantidade de código a ser implantada além de aumentar a segurança e performance da aplicação, diminui o tempo de inicialização \cite{madhavapeddy2014unikernels}.

Além de se aproveitar do rápido \textit{boot} do unikernel, o método \textit{SEUSS} armazena em memória cache o estado da execução do unikernel em várias etapas de invocação de uma função, esses \textit{snapshots} são chamados de UC (\textit{Unikernel Context}). O objetivo dessa estratégia é evitar que, durante o processo de criação de novas instâncias, seja necessário realizar as mesmas etapas de inicialização.

O método \textit{SEUSS} considera que quando não existe nenhuma réplica da função disponível, a invocação de uma função em uma plataforma \textit{serverless} pode seguir por três caminhos: \textit{coldest path}, \textit{warm path} e \textit{hot path}. Para o \textit{SEUSS}, quando a invocação inclui o \textit{coldest path}, é necessário realizar todos os passos de inicialização da aplicação discutidos na seção \ref{section:introduction}, incluindo construir o ambiente de execução isolada. Já no \textit{warm path}, o método considera que a execução da função será atrasada pelo fato que o código fonte da função ainda não foi carregado e compilado. E finalmente, no \textit{hot path}, existe um UC da função em que todas as etapas de \textit{startup} foram realizadas, sendo necessário apenas restaurar o \textit{snapshot} para processar a requisição.

Para evitar o atraso de inicialização do ambiente de execução isolada e da \textit{runtime}, o \textit{SEUSS} propõe que quando um novo nó de computação se junta aos recursos computacionais da plataforma, um processo de inicialização deve ser realizado. Por exemplo, para cada \textit{runtime} suportada pela plataforma, um binário do unikernel deve ser executado contendo o código da \textit{runtime} e um script de invocação. O script é responsável por criar um servidor HTTP, e importar o código fonte da função quando necessário. Após o processo de inicialização, o \textit{snapshot} é realizado criando um UC. Como neste momento de \textit{snapshot} o UC ainda não carregou nenhuma função, então pode ser utilizado para criar réplicas de funções que compartilham a mesma \textit{runtime}. Com essa estratégia, uma função pode trilhar apenas o \textit{warm} e \textit{hot path} durante um evento de \textit{cold-start}.

Apesar da exclusão do \textit{coldest path}, lidar com o \textit{warm path} também é importante. O processo de importar, compilar o código fonte de uma função e suas dependências pode durar tanto quanto iniciar um contêiner ou VM. Para lidar com este problema, o \textit{SEUSS} define que sempre que uma função é invocada, e existe apenas o UC com a \textit{runtime} genérica disponível, o método \textit{SEUSS} deve restaurar o UC e realizar todo o processo de inicialização da função. Dessa forma, quando a função fica pronta para processar requisições, o método realiza o \textit{snapshot} da função, que irá armazenar em cache um UC que evita todas as etapas de inicialização.

Com a união de todas as estratégias para lidar com o \textit{coldest} e \textit{warm path}, sempre que for necessário criar uma nova réplica de uma função, o \textit{SEUSS} pode verificar qual o UC com o estado de inicialização mais avançado disponível em cache, e em seguida restaurar o estado do UC para implantar a nova instância da função evitando etapas de \textit{startup}.

\subsection{Prebaking vs SEUSS}

A principal diferença entre a estratégia de \textit{Prebaking} e o \textit{SEUSS} é que \textit{Prebaking} realiza apenas um \textit{snapshot} da aplicação. A técnica define que o \textit{snapshot} deve ser gerado quando a função completa todo o processo de inicialização. Além disso, o \textit{snapshot} deve ser feito no momento em que a função é implantada na plataforma, e não no momento de sua invocação como proposto pelo método \textit{SEUSS}.

O \textit{SEUSS} optou por essa estratégia para reduzir a quantidade de memória utilizada pelas aplicações \textit{serverless}. Realizar \textit{snapshots} em diferentes momentos do ciclo de vida de uma função, facilita o compartilhamento de páginas de memória entre UCs por meio do mecanismo de \textit{copy-on-write}.

Por outro lado, a consequência de gerar o \textit{snapshot} no momento da invocação da função, é que as funções ainda terão que pagar o preço de carregar e compilar o código fonte. Apesar de salvar um UC após esse processo, esse evento de \textit{cold-start} irá acontecer sempre que não houver uma réplica disponível, ou um \textit{snapshot} com a função pronta. Além disso, funções que possuem muitas dependências podem aumentar o tempo de inicialização para casa de segundos.

Outra diferença entre as duas estratégias é que o método \textit{SEUSS} utiliza unikernels como tecnologia para isolar a execução das funções. Porém, unikernels precisam de uma implementação especializada de sistema operacional, isso significa que os nós de computação da plataforma não serão hospedados por um sistema operacional Linux. Apesar de todas as vantagens do unikernel, a consequência da sua utilização pode tornar mais complexa a integração do método com plataformas \textit{serverless}, aumentando a complexidade e custo de sua adoção.

\section{Metodologia}

Para garantir a generalização dos resultados, avaliamos o desempenho das estratégias para acelerar o \textit{cold-start} de uma função tipicamente \textit{serverless}: \textit{Markdown} \cite{silva2020prebaking, wang2019replayable, shahrad2019architectural}, e também de funções sintéticas como \textit{NoOp} e \textit{Big} \cite{cadden2020seuss, silva2020prebaking}. Todas as funções foram implementadas utilizando NodeJS\footnote{\url{https://github.com/paulofelipefeitosa/serverless-handlers/tree/master/functions/nodejs/}}\footnote{\url{https://github.com/paulofelipefeitosa/serverless-handlers/tree/master/seuss/functions/}}, que uma das linguagens mais utilizadas por aplicações \textit{serverless} \cite{passwater2018serverless}.

A função \textit{Markdown} converte documentos escritos em \textit{markdown} para páginas HTML. A responsabilidade de conversão é delegada para biblioteca \textit{showdown}\footnote{\url{http://showdownjs.com/}}. Para eliminar possíveis ruídos experimentais advindos da comunicação em rede, ao ser invocada, a função lê um \textit{markdown}\footnote{\url{https://github.com/PrincetonUniversity/openpiton/blob/openpiton/README.md}} armazenado no disco local, que em seguida deve ser convertido em HTML.

A função sintética \textit{NoOp} não possui nenhuma lógica de negócio, e nem carrega dependências, portanto consideramos essa função a base para o tempo de inicialização de qualquer função \textit{serverless}. Ao ser invocada, a função \textit{NoOp} sempre envia uma resposta com o corpo vazio e status de sucesso.

Por outro lado, apesar da função sintética Big também não implementar lógica de negócio, durante o seu \textit{startup}, 41 MB de dependências são carregadas. A importação das dependências acontece de maneira forçada durante a execução da função Big. Essa característica permitirá entender o comportamento das estratégias de \textit{snapshot} quando a função carrega uma grande quantidade de dependências.

Para ambas estratégias repetimos a execução de cada função 100 vezes. Em cada execução, foram disparados 100 eventos sequenciais a uma taxa constante requisitando a execução das funções. Durante cada evento, medimos a latência de resposta e o tempo de serviço necessário para as funções processarem os eventos. 

\begin{figure}[!h]
    \centering
    \includegraphics[scale=0.44]{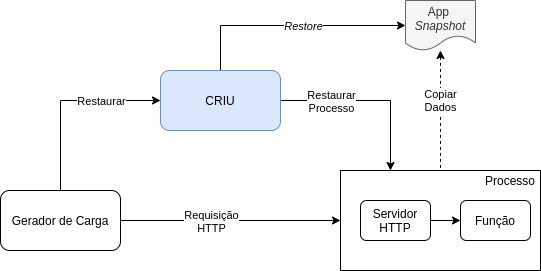}
    \caption{Componentes e fluxo de comunicação envolvidos na avaliação da técnica de \textit{Prebaking}.}
    \label{fig:prebaking_diagram}
\end{figure}

A Figura \ref{fig:prebaking_diagram} ilustra os componentes envolvidos na avaliação da técnica de \textit{Prebaking}, que inclui: um gerador de carga; o CRIU; e o \textit{snapshot} da função. O gerador de carga é o componente responsável por requisitar a restauração da função e, em seguida, disparar uma sequência de requisições para serem processadas. Durante a inicialização e o processamento das requisições, o gerador de carga é responsável por medir o tempo que leva para a função aceitar a primeira requisição, assim como o tempo de serviço das requisições. O CRIU é a ferramenta responsável por recuperar o processo da função utilizando o seu \textit{snapshot}, sendo o \textit{snapshot} uma imagem do processo da função previamente salva em disco por meio da técnica de \textit{Prebaking}.

\begin{figure}[!h]
    \centering
    \includegraphics[scale=0.38]{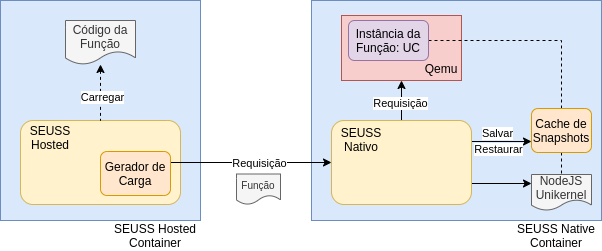}
    \caption{Componentes e fluxo de comunicação envolvidos na avaliação do método \textit{SEUSS}.}
    \label{fig:seuss_diagram}
\end{figure}

Como exposto pela Figura \ref{fig:seuss_diagram}, a avaliação do método \textit{SEUSS} é composta por dois contêineres. O primeiro contêiner executa o componente \textit{hosted} do \textit{SEUSS} que utiliza bibliotecas do sistema operacional de propósito geral. Tal característica, permite a comunicação das instâncias do \textit{SEUSS} com componentes externos. O segundo contêiner executa o componente nativo do \textit{SEUSS}. Esse componente permite a execução de aplicações que se comunicam diretamente com as interfaces de hardware.

O \textit{SEUSS Hosted} já implementa um gerador de carga para as aplicações, que além de permitir o teste das funções, também computa o tempo de inicialização da função e o tempo de serviço de cada requisição. Durante a execução do teste, o gerador carrega e envia o código da função para o \textit{SEUSS} Nativo requisitando a sua execução. Por sua vez, o \textit{SEUSS} Nativo é o responsável por checar se existe algum UC em cache para a função, ou checar se será necessário criar um novo UC a partir do unikernel de NodeJS que utiliza o \textit{Rumprun}\footnote{\url{https://github.com/rumpkernel/rumprun}}. Com o UC em mãos, o \textit{SEUSS} Nativo requisita a execução do binário com o emulador de hardware \textit{QEMU}\footnote{\url{https://www.qemu.org/}}.

Todas as execuções para ambas estratégias foram realizadas em um computador Intel(R) Core(TM) i7-8550U 1.80GHz de 8 núcleos, com 16GB de memória RAM executando o Ubuntu 18.04.5 LTS com o kernel Linux v4.15.0-142-generic-x86\_64. Utilizamos como \textit{runtime} a versão v4.3.0 do NodeJS com NPM v2.14.12, e o CRIU v3.6.0. Antes de cada execução, o gerador e os ambientes de execução de \textit{Prebaking} e \textit{SEUSS} foram reiniciados.

\section{Resultados}

Nesta seção apresentamos os resultados das execuções das técnicas de \textit{Prebaking} e \textit{SEUSS}. Primeiro comparamos o desempenho de ambas as técnicas para inicializar a execução das funções, e em seguida avaliamos o desempenho do tempo de serviço das funções para processar a primeira e as demais requisições.

\subsection{Cold-start}

\begin{figure*}[!h]
    \centering
    \includegraphics[scale=0.6]{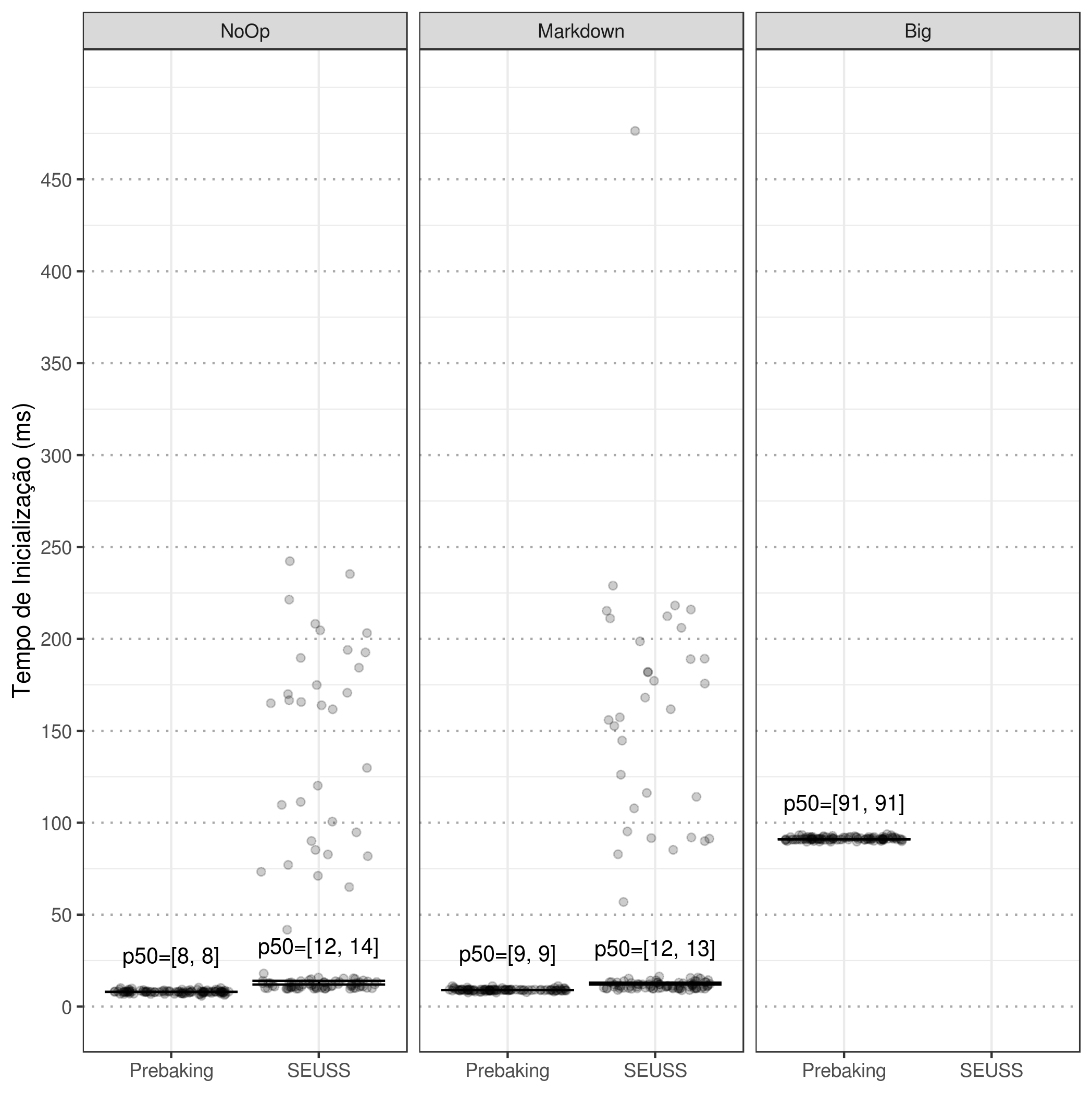}
    \caption{Gráfico de pontos do tempo de inicialização de cada função versus a técnica empregada para reduzir o \textit{cold-start}. As barras de erro representam o intervalo da mediana com confiança de 95\% utilizando \textit{bootstrap}.}
    \label{fig:startup_latency}
\end{figure*}

A Figura \ref{fig:startup_latency} compara o tempo de inicialização em milissegundos das funções \textit{NoOp}, \textit{Markdown} e \textit{Big} quando empregadas as técnicas de \textit{Prebaking} e \textit{SEUSS}. As barras de erro representam o intervalo da mediana com confiança estatística de 95\% utilizando \textit{bootstrap} \cite{efron1994introduction}. Pode-se notar que a função \textit{Big} não possui observações de tempo de inicialização quando executada com o \textit{SEUSS}, isto porque nenhuma execução da função foi bem sucedida. Por outro lado, falhas nas execuções de \textit{NoOp} e \textit{Markdown} com o \textit{SEUSS} também ocorreram, contudo a reinicialização da execução mitigou esse problema. A razão das falhas na execução ainda é desconhecida e precisa ser investigada.

Apesar da frequente ocorrência de valores distantes da mediana para as execuções com o \textit{SEUSS}, é difícil concluir de maneira visual qual estratégia acelera mais o \textit{cold-start} em cada cenário. Porém, o intervalo de confiança disjunto das medianas parece apontar para um melhor desempenho de \textit{Prebaking}. Para confirmar que as medianas não são iguais, realizamos o teste não-paramétrico de \textit{Wilcoxon-Mann-Whitney}. E os testes confirmaram, existe uma forte evidência de que \textit{Prebaking} apresenta um melhor desempenho para diminuir o tempo de inicialização das funções \textit{NoOp} e \textit{Markdown}. Para a função \textit{NoOp} a diferença de desempenho mediana ficou entre [4, 6] milissegundos, representando uma diferença percentual entre 33\% e 42\% à favor da técnica de \textit{Prebaking}. Já para a função \textit{Markdown} a diferença ficou entre [3, 4] milissegundos, uma diferença percentual entre 25\% e 30\%. 

Como \textit{NoOp} é uma função muito simples, acaba se tornando a linha de base para avaliar o desempenho de técnicas para acelerar o \textit{cold-start}. É possível observar que para \textit{Prebaking} à medida que a função se torna mais complexa (com mais dependências), o tempo para sua inicialização aumenta. Por outro lado, não foi possível observar o mesmo comportamento com o \textit{SEUSS}. Existe uma intersecção no intervalo de confiança do tempo de inicialização mediano das funções, isto pode significar que o desempenho de \textit{startup} do \textit{SEUSS} não seja afetado pela complexidade da função. Uma avaliação com uma função mais complexa do que \textit{Markdown} poderia concluir essa hipótese.

\subsection{Tempo de serviço}

Uma outra métrica que também pode sofrer distorções por causa do \textit{cold-start} é o tempo de serviço da primeira requisição. Este fenômeno acontece devido a mecanismos de compilação \textit{just-in-time} (JIT) que são amplamente oferecidos por linguagens de programação gerenciadas por ambientes de execução virtualizados. 

\begin{figure*}[!h]
    \centering
    \includegraphics[scale=0.6]{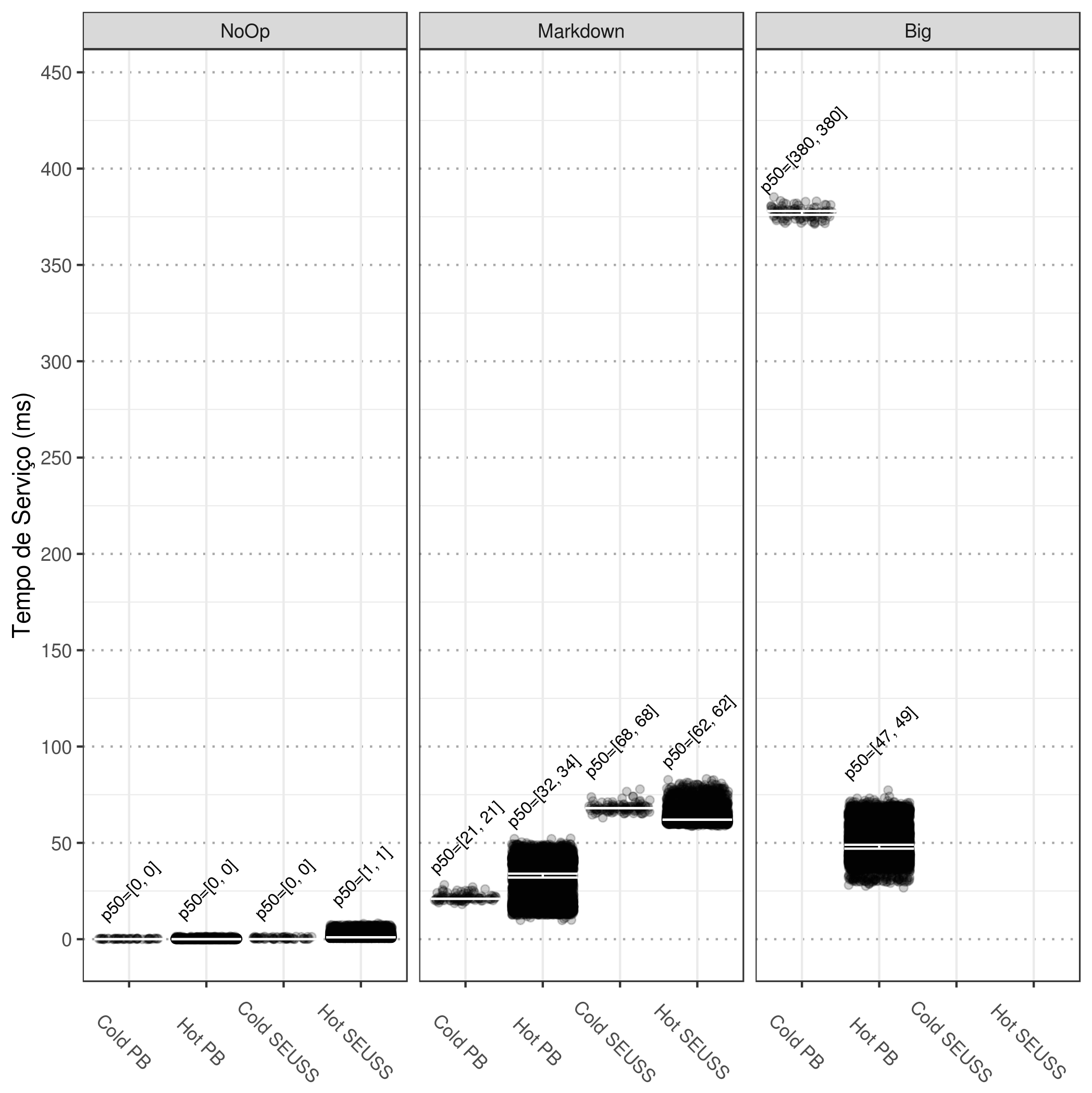}
    \caption{Gráfico de pontos do tempo de serviço das requisições disparadas para cada função versus a técnica empregada para reduzir o \textit{cold-start} junto com o cenário de disparo. As barras de erro representam o intervalo da mediana com confiança de 95\% utilizando \textit{bootstrap}.}
    \label{fig:service_time}
\end{figure*}

A Figura \ref{fig:service_time} apresenta o desempenho do tempo de serviço das funções para ambas as técnicas. Além disso, para entender se o fenômeno de JIT também afeta o desempenho de \textit{Prebaking} e \textit{SEUSS}, classificamos o tempo de serviço em duas formas: 1) \textit{Cold}, significando o tempo de serviço da primeira requisição processada pela função; 2) \textit{Hot}, representando o tempo de serviço das requisições processadas após a primeira. Dessa forma, é possível saber se houve alguma penalidade no tempo de serviço das requisições \textit{Cold} quando comparadas com a \textit{Hot}.

Como \textit{NoOp} é uma função simples, que não possui dependências, não esperávamos que o JIT afetasse o desempenho do tempo de serviço da primeira requisição. Podemos observar que para a técnica de \textit{Prebaking}, em ambos os cenários, a mediana do tempo de serviço possui a mesma tendência. Contudo, para o \textit{SEUSS} no cenário \textit{Hot} a mediana apresenta um leve aumento de 1 ms, e também apresenta uma maior variabilidade de valores, indicando que o cenário \textit{Hot} teve desempenho pior em relação ao cenário \textit{Cold}.

Como a função \textit{Markdown} é mais complexa, pois possui uma dependência externa e carrega um arquivo de 28.1 KB durante a sua inicialização, é natural que a primeira requisição tenha um tempo de serviço maior do que o das requisições posteriores. Contudo, a técnica de \textit{Prebaking} possui um mecanismo de aquecimento que força a execução da função antes de realizar o \textit{snapshot} do processo. Dessa forma, esperávamos que o desempenho do tempo de serviço em ambos os cenários fossem estatisticamente iguais. Mas, como é possível notar, a execução com \textit{Prebaking} no cenário Cold apresenta um melhor desempenho quando comparado com o cenário \textit{Hot}, uma redução de 34\% no tempo de processamento. Acreditamos que a ocorrência desse fenômeno pode estar associada a pausas para coleta de lixo \cite{quaresma2020controlling}.

Por outro lado, esperávamos que o desempenho do tempo de serviço da função \textit{Markdown} com o \textit{SEUSS} no cenário \textit{Cold} fosse afetado pelo JIT. Nossa hipótese era que o fato da técnica realizar o cache das funções por demanda, seria necessário compilar o código no momento de processar a primeira requisição. E os resultados apontam para uma forte evidência dessa hipótese, a mediana do cenário \textit{Cold} é de aproximadamente 68 ms enquanto que no cenário \textit{Hot} é de aproximadamente 62 ms, uma diferença de 9,6\%.

Assim como para a função \textit{Markdown}, esperávamos que o tempo de serviço da função \textit{Big} utilizando \textit{Prebaking} fossem estatisticamente iguais em ambos os cenários. Contudo, o tempo de serviço foi significativamente maior no cenário \textit{Cold}, a diferença entre as medianas chega a ser 7x maior. A ocorrência desse fenômeno pode estar associada ao funcionamento interno do NodeJS v4.3.0, pois o mesmo fenômeno não foi observado durante a avaliação de outras \textit{runtimes}, como Java \cite{silva2020prebaking}.

Uma outra curiosidade é que o desempenho do tempo de serviço do \textit{SEUSS} para a função \textit{Markdown} foi significativamente maior do que o tempo de serviço de \textit{Prebaking}. Como essa métrica mensura o tempo que leva para a função processar a requisição, esperávamos que os valores para fossem estatisticamente iguais. A causa dessa diferença ainda é desconhecida, porém temos como prováveis causas as seguintes hipóteses: 

\begin{enumerate}
    \item a forma como o SEUSS importa o código da função não permite o reuso de variáveis, fazendo com que a importação de dependências necessárias a inicialização sejam re-executadas a cada invocação da função;
    \item o fato do SEUSS ser executado pelo \textit{QEMU}, um ambiente virtualizado de hardware, e a solução ter sido empacotada por um container, podem ter contribuído para a degradação da performance da técnica \cite{bellard2005qemu, casalicchio2017measuring, raho2015kvm};
    \item o ambiente onde o SEUSS foi avaliado \cite{cadden2020seuss} possui uma quantidade e qualidade superior aos recursos computacionais disponíveis em nosso ambiente.
\end{enumerate}

\section{Conclusão}
Plataformas \textit{serverless} sofrem de um problema conhecido como \textit{cold-start}. Esse problema se apresenta como um atraso no processamento de requisições devido a espera pelo escalonamento de recursos para hospedar a execução das aplicações. Diversos trabalhos já foram publicados com propostas de soluções para o \textit{cold-start}. Dentre estes trabalhos se destacam as estratégias que utilizam o mecanismo de \textit{snapshot}, que evita a re-computação de etapas de inicialização comum entre funções. Contudo, existe uma carência de trabalhos que discutam as vantagens e desvantagens das soluções que adotam o mecanismo de \textit{snapshot}.

Para preencher essa lacuna, realizamos a comparação de desempenho de duas estratégias que adotam o mecanismo de \textit{snapshot}: 1) a técnica de \textit{Prebaking}; 2) o método \textit{SEUSS}. A principal diferença entre as estratégias é que o \textit{SEUSS} utiliza unikernel para isolar a execução das funções, enquanto \textit{Prebaking} faz o uso de sistemas operacionais de propósito geral. Conseguimos avaliar o desempenho das técnicas levando em consideração o tempo de inicialização e o tempo de serviço de duas funções: 1) \textit{NoOp} e 2) \textit{Markdown}.

Sobre a ótica do tempo de inicialização das funções, que indica o atraso esperado para o atendimento das requisições, a técnica de \textit{Prebaking} apresentou um desempenho 33\% melhor do que o método \textit{SEUSS} para a função \textit{NoOp}, e um desempenho 25\% melhor para a função \textit{Markdown}. Quando analisamos possíveis penalidades adicionais para processar as requisições dos clientes, o método \textit{SEUSS} levou cerca de 9,6\% a mais de tempo para processar a primeira requisição da função \textit{Markdown}, enquanto que a técnica de \textit{Prebaking} reduziu o tempo de processamento da primeira requisição em cerca de 34\%. Esse resultado evidenciou a importância da adoção de mecanismos de \textit{warmup} para lidar com o impacto do \textit{cold-start} no tempo serviço da primeira requisição.

Durante os experimentos com o \textit{SEUSS}, diversas execuções não foram bem sucedidas, e para viabilizar nossas análises foi necessário realizar a reexecução dos casos falhos. Contudo, para a função Big, todas as execuções falharam. Assim, inviabilizando a comparação de desempenho das técnicas com esta função. Dessa forma, como um possível trabalho futuro, pretendemos investigar a causa das falhas para permitir a avaliação de outras funções tão complexas quanto a função Big, e assim expandir a comparação de desempenho das duas técnicas. Também planejamos investigar a causa da degradação de desempenho do tempo de serviços das funções executadas com o método \textit{SEUSS}. Por fim, queremos entender a não efetividade do mecanismo de \textit{warmup} da técnica de \textit{Prebaking} para a função Big implementada em NodeJS.

\bibliographystyle{ACM-Reference-Format}
\bibliography{sample-base}

\end{document}